\documentclass[screen,sigconf]{acmart}

\AtBeginDocument{%
  \providecommand\BibTeX{{%
    \normalfont B\kern-0.5em{\scshape i\kern-0.25em b}\kern-0.8em\TeX}}}

\copyrightyear{2024}
\acmYear{2024}
\setcopyright{rightsretained}
\acmConference[ASSETS '24]{The 26th International ACM SIGACCESS Conference on Computers and Accessibility}{October 27--30, 2024}{St. John's, NL, Canada}
\acmBooktitle{The 26th International ACM SIGACCESS Conference on Computers and Accessibility (ASSETS '24), October 27--30, 2024, St. John's, NL, Canada}
\acmDOI{10.1145/3663548.3688487}
\acmISBN{979-8-4007-0677-6/24/10}

\usepackage{pifont}
\usepackage{enumitem}

\begin{document}

\title{StereoMath: An Accessible and Musical Equation Editor}

\author{Kenneth Ge}
\orcid{0009-0000-5044-4433}
\affiliation{%
  \institution{Assistivity}
  \city{Edgemont}
  \state{NY}
  \country{USA}
  \postcode{10583}
}
\email{kennethkouge@gmail.com}

\author{JooYoung Seo}
\orcid{0000-0002-4064-6012}
\affiliation{%
 \department{School of Information Sciences}
  \institution{University of Illinois Urbana-Champaign}
  \city{Champaign}
  \state{Illinois}
  \country{USA}
  \postcode{61820}
  }
\email{jseo1005@illinois.edu}


\begin{abstract}

For blind and low-vision (BLV) individuals, digital math communication is uniquely difficult due to the lack of accessible tools. Currently, the state of the art is either code-based, like LaTeX, or WYSIWYG, like visual editors. However, both paradigms view math communication as primarily a visual typesetting problem, and may be accessible but difficult to use. In this paper, we present an equation editor that is built from the ground up with BLV accessibility in mind. Specifically, we notice that two of the biggest barriers with current technology are the high cognitive load and the lack of spatial relationships. Thus, we build an editor that uses spatial audio cues, muscle memory, tones, and more intuitive navigation to properly contextualize math equations. We discuss how this new paradigm can enable new levels of math communication, engagement, and literacy. Finally, we discuss natural next steps. 
\end{abstract}



\received{July 2024}
\received[revised]{August 2024}
\received[accepted]{August 2024}

\maketitle

\section{Introduction}
\label{sec:introduction}

Beyond the intrinsic difficulty of math lies a pernicious barrier: understanding abstract notation. Unintuitive abstractions are known to impede learning and reduce performance \cite{widada_overcoming_2020,wason_natural_1971,koedinger_real_2004}. The proliferation of graphical equation editors, live TeX typesetters, and sophisticated autocomplete suggests this is a huge problem: Knauff \cite{knauff_efficiency_2014} observed that "on most measures, expert LaTeX users performed even worse than novice Word users." 

However, the existing tools for blind and low-vision (BLV) individuals rely largely on verbal readalouds, which users have described as long and confusing \cite{da_paixao_silva_how_2017}, and coding languages such as LaTeX, which are unintuitive and require their own domain knowledge \cite{seo_latex_2019}. As Monzoor \cite{manzoor_alap_2019} writes, mathematics is "a two dimensional, non-linear language that requires equally complex representation." This information loss is one of the biggest challenges for BLV users \cite{stoger_accessing_2015}, especially in Braille and specialized Nemeth Code \cite{sauer_mathematics_2020}. All of this extra burden results in editors that have high cognitive load and are difficult to navigate. On complex interfaces, blind users spend over 30\% longer on navigation than sighted users \cite{lazar_what_2007}. In math, this burden on working memory may be especially harmful \cite{engle_working_2002,ashcraft_relationships_2001}. 

Even in math problems that are not obviously visual in nature, empirical tests \cite{van_garderen_visualspatial_2003,vale_importance_2017} and brain scans \cite{knauff_spatial_2002} validate the importance of spatial reasoning. Like sighted people, blind individuals show increased activation in the visual cortex \cite{kanjlia_absence_2016,amalric_role_2018}, reason about complex spatial representations \cite{tinti_visual_2006}, and use a spatial number line \cite{robertson_problem_2016, castronovo_semantic_2007}. BLV users also prefer and perform better with visual-spatial aides, compared to verbal-only channels \cite{szpiro_how_2016}.

To address these issues, we propose StereoMath, a web-based accessible equation editor. It provides: 1) representations of spatial relationships, 2) a higher information density, and 3) a more intuitive sense of nestedness between equation elements. With a new interface for both reading and writing, our work aims to help individuals at all levels, from beginners hoping to better understand written equations, to mathematicians seeking a more efficient tool to transcribe with. We focused on the following research questions:
\begin{itemize}[topsep=0pt]
    \item How can an editor that incorporates spatial and ownership relations improve depth of understanding?
    \item How can a sonified editor enhance the user experience and help blind individuals engage with STEM?
    \item How can an editor that is optimized for cognitive load increase productivity?
\end{itemize}

\section{Related Work}
\label{sec:related_work}

Most related works in this space rely on LaTeX coding. For example, ALAP \cite{manzoor_alap_2019} focuses on improving the LaTeX debugging experience, and tools like AROW \cite{seo_latex_2019} and WriteR \cite{godfrey_ajrgodfreywriter_2024} offer easier document layout editing, but still rely on LaTeX coding for equations. However, LaTeX is known to have a steep learning curve, with extra challenges for BLV users \cite{seo_latex_2019}. 

The most dominant paradigm for reading formulae is in a linear fashion, such as in ChromeVox and NVDA. Unfortunately, this linear representation results in "fewer possibilities to 'navigate' between lines, besides the need to memorize terms and partial results" \cite{da_paixao_silva_content_based_2018,da_paixao_silva_how_2017}. With Nemeth Code Braille-based readers, there is an added cost of learning Braille and purchasing a refreshable display. 

Some tools, such as JAWS MathViewer and ChromeVox-NavMatBr \cite{da_paixao_silva_content_based_2018}, organize mathematical expressions into a tree, enabling faster and more comprehensible navigation \cite{da_paixao_silva_how_2017}.Unfortunately, JAWS does not adhere to the MathSpeak protocol, resulting in ambiguous readalouds: e.g., ``$\sqrt{3i}-2i$'' and ``$\sqrt{3i-2i}$'' result in the same output \cite{noauthor_jaws_nodate,noauthor_jaws_nodate_1}. 

Luckily, work in other domains has successfully overcome similar challenges, by using sonification \cite{holloway_infosonics_2022,barrass_using_1999} (e.g., in SAS Graphics Accelerator \cite{noauthor_sas_nodate} and EmacsSpeak \cite{raman_emacspeak___speech_1996}), as well as employing modality switching \cite{paivio_dual_1991}, which has been shown to reduce cognitive load \cite{cao_modality_2009,klatzky_cognitive_2006,brunken_assessment_2004}. 

To our best knowledge, this work is the first equation editor built to mirror the intuitive nature of sighted equivalents. The only web-based alternative, Desmos \cite{noauthor_desmos_nodate}, is closed-source and lacks LaTeX code generation. 

\section{Design Procedures and Goals}
\label{sec:design_procedures_and_goals}

\subsection{Design Procedures}
\label{subsec:design_procedures}

We used a mixed-ability co-design process, following the interdependence framework \cite{bennettInterdependenceFrameAssistive2018}. Our team consists of two researchers, one blind researcher (BR) and one sighted researcher (SR). BR has been using screen readers for over 23 years, and LaTeX for over 10. Meanwhile, SR has been using LaTeX for 4 years.  

Our work consisted of a series of weekly testing sessions since February through July, 2024. Each week, we would test a prototype, brainstorm potential new ideas and features, and then SR would implement these before regrouping to test. Having a mixed-ability team helped us identify effective and novel solutions, while having a sighted member enabled us to add features from sighted WYSIWYG editors missing in the BLV space. This project began as a simple tree-based extension of autocomplete, but quickly grew to a much larger project through our co-design sessions. Much of our inspiration came from our combined background in audio/music production, web development, and game development. 

\subsection{Design Goals}
\label{sec:design_goals}

Inspired by our co-design sessions, we established the following design goals:

\textbf{DG1: The editor should convey spatial information through binaural/stereo audio and pitch.} The popular linear navigation style fails to differentiate elements that are on the lefthand or righthand side of an equation or operator. They also fail to convey when two elements are in the same horizontal space, such as the numerator and denominator of a fraction, or the same vertical space, such as a series of plus operators. Although temporal information serves as a proxy for left/rightedness, this puts a strain on working memory, whereas these clues are conveyed subconsciously to sighted users.

\noindent\textbf{DG2: The editor should encode spatial relationships through muscle memory and spatial keyboard navigation.} Stereo audio is limited in only conveying left/right relationships, since back/front and up/down audio relies on spectral cues that are specific to the shape of an individual's ears \cite{musicant_influence_1984}. Instead, we use the keyboard to encode a sort of spatial muscle memory.

\noindent\textbf{DG3: The editor should utilize earcons to shift information out of the verbal modality.} The working memory can only hold so many terms at once \cite{miller_magical_1956}. Thus, we replace verbal elements that are not particularly meaningful (which we call "verbal filler"), with nonverbal cues. This allows us to increase the number of non-filler terms in working memory. For example, terms like "lparen" and "rparen" would be replaced with brief tones. 

\noindent\textbf{DG4: The editor should be able to precisely and intuitively contextualize equation fragments, and convey a sense of "ownership" between elements of an equation.} One example use case is a way to quickly gauge that an element is under a square root, or in the numerator of a fraction. We want an intuitive way of expressing difficult relationships such as nested equation fragments and recursive fractions. 

\noindent \textbf{DG5: The final result should be customizable.} For example, users can import their own keyboard layout for spatial navigation, or adjust stereo sound parameters to accommodate hearing impairments

\section{System Implementation}
\label{sec:system_implementation}

\subsection{Overview}
\label{subsec:system_overview}

\begin{figure}
    \centering
    \includegraphics[height=0.25\textwidth]{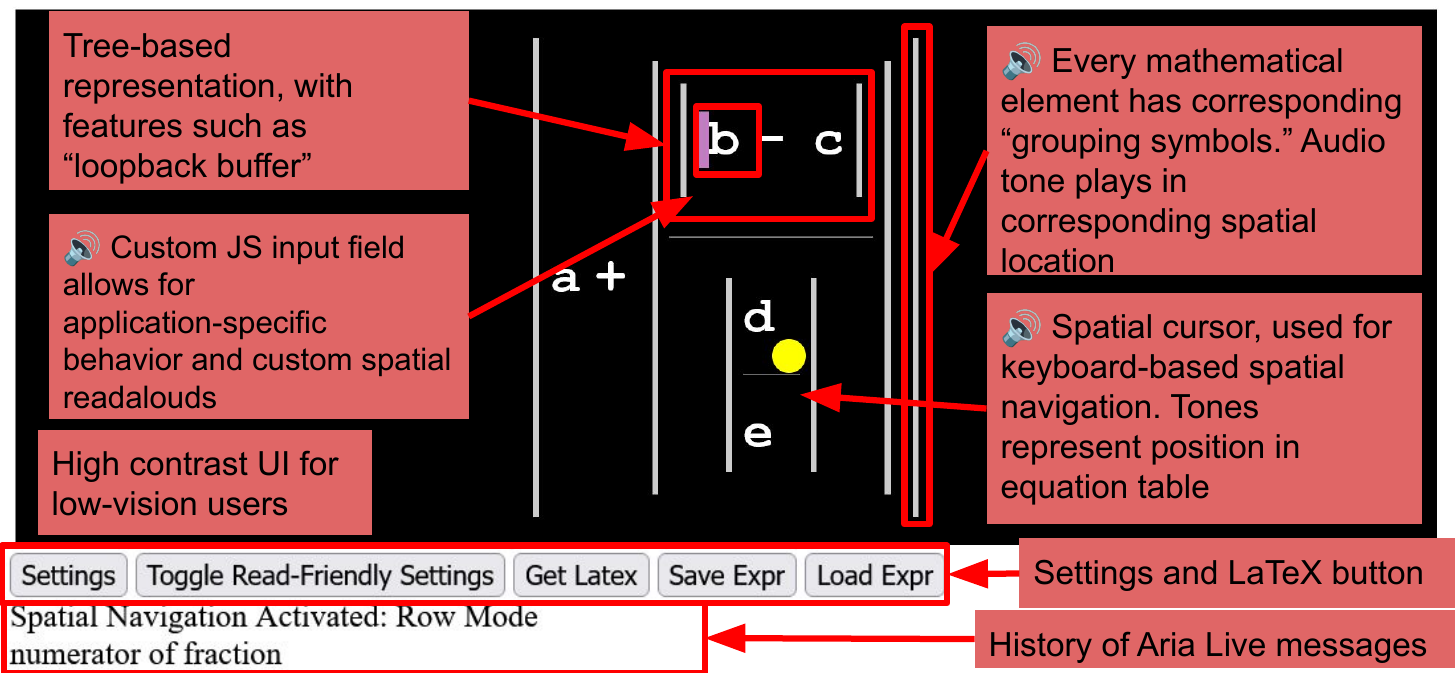}
   \Description[A diagram detailing the different elements of the equation editor]{A screenshot of the equation editor, along with arrows and boxes highlighting different elements. Elements are as follows: A box highlighting a node, with caption "tree-based representation, with features such as loopback buffer"; a box highlighting an input field with caption "custom JS input field allows for application-specific behavior and custom spatial readalouds"; a box highlighting the vertical parentheses bars with caption "Every mathematical element has corresponding “grouping symbols.” Audio tone plays in corresponding spatial location"; a box highlighting the spatial cursor, with caption "Spatial cursor, used for keyboard-based spatial navigation. Tones represent position in equation table". Finally, the image also highlights the Settings and LaTeX generation buttons, as well as the live aria messages}
   \caption{System Overview Diagram}
   \label{fig:overview}
\end{figure}

StereoMath is a web-based equation editor, built from the ground-up in Node.js and HTML/CSS/JS, using the backend server to generate custom spatial text to speech. Figure~\ref{fig:overview} provides a high-level overview of the interactive elements. 

We conceptualize an equation in two ways: A) as an equation tree, with operators acting as internal nodes, and variables/numbers representing leaves, and B) as a table/grid, where rendering the equation out in 2D is equivalent to laying elements out in their respective rows and columns. This novel table-based conceptualization is useful for spatial navigation. 

The following are some important features, aligning with our design goals.

\noindent\textbf{Spatial Sound (addresses DG1)} Arguably some of our most important innovations come from our spatial sound and spatial navigation. Here, every time the user interacts in a way that produces sound (e.g., navigation or typing), we modify the audio to match the spatial left-right location in the equation's final rendering. We also change the pitch to match the vertical up/down location. In our co-design sessions, this improved ease of comprehension, and provided a novel experience that was previously not commonly found elsewhere. 

\noindent\textbf{Spatial Keyboard Navigation (addresses DG2)} We wanted to reinforce the spatial relationships using a tactile modality and muscle memory, without any new equipment. Thus, we designed a new navigation mode called "spatial nav mode," which uses the user's own keyboard as a clickable map of the layout. Upon typing a key, we indicate which element has been selected using a series of musical tones, as well as a verbal readaloud. We also added a virtual cursor to traverse the equation grid using the arrow keys. More details in Section \ref{subsec:system_architecture}.

\noindent\textbf{Earcons (addresses DG3)} To increase the density of information, we sonify many elements. For example, the "pseudo-parentheses" are sonified with opening and closing earcons. This allows us to avoid the longer "Frac" and "EndFrac" that are suggested by the MathSpeak protocol. We also use earcons to indicate when the user has reached the end of text in equation navigation style (avoiding the long "end of text" that screen readers will play by default), when the user has entered an editable text field, and when the user has navigated onto a node element (e.g., a "plus" element). 

\noindent\textbf{Tree Navigation (addresses DG4)} Implementing the equation as a tree required a few important considerations. First, we introduce two different navigation styles, "linear" and "equation." The linear style corresponds to a linear scan through the entire text of the equation, and behaves similar to moving the caret across each character one-by-one. Through our co-design sessions, we found that this was more intuitive for first-time blind users, because of its ubiquity elsewhere. However, certain advanced functionality was more intuitive in "equation mode," which mimics the experience of having distinct text fields in WYSIWYG editors. 

Second, we introduce "loopback buffers," which allow the user to read out any selected node (or the entire equation) in both MathSpeak and "IntuitiveSpeak." IntuitiveSpeak is less precise, but shorter than MathSpeak, and has a one-to-one correspondence with the elements on the page. E.g., MathSpeak might produce "Base Frac x Over y EndFrac," while IntuitiveSpeak would produce "x over y."

Third, we include "pseudo-parentheses" to mark the start and end of each node. These are not included in the final LaTeX rendering, but reduce ambiguity by contextualing any nested elements. 

Finally, we introduce code folding, which collapses operators to simplify the expression. This, plus the above features, combines together into a "read friendly mode" that the user can toggle.

\noindent\textbf{High-Contrast, Settings, and Extensibility (addresses DG5)} We introduce a high-contrast UI for individuals who may have limited vision. We also add the ability to modify settings such as the verbosity of the editor, the navigation style, and turning off spatial audio (e.g. for those with unilateral hearing loss). 

Finally, we add the ability to generate LaTeX code, and also load and save equations. 

\subsection{Architecture}
\label{subsec:system_architecture}

\begin{figure}
    \centering
    \includegraphics[height=0.3\textwidth]{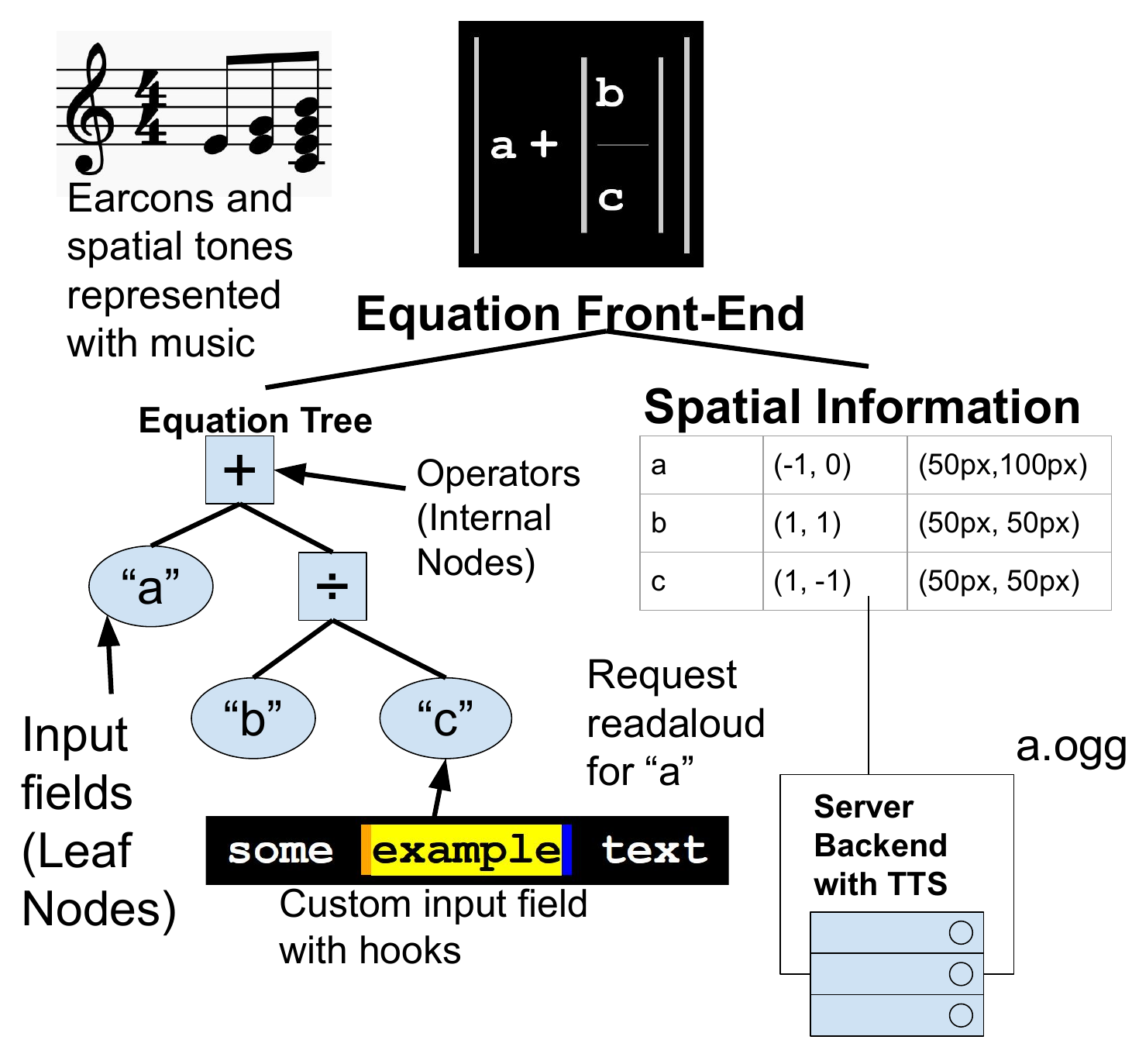}
   \Description[Architecture overview, with trees and spatial info]{A diagram featuring the internal equation tree on the left side, and the spatial information on the right side, which is connected to a custom TTS on the server-side. The internal nodes on the equation tree are labeled as operators, and the leaf nodes are input fields. It also shows an example custom input field, and an example series of musical chords representing an earcon.}
   \caption{Architecture overview, showing the two representations (tree-based and spatial/geometric), as well as the custom backend TTS. }
   \label{fig:architecture}
\end{figure}

On the server-side, our biggest architectural hurdle was implementing our cusotm TTS. We piped announcements to the server, generated raw PCA audio on the server using Piper TTS \cite{noauthor_rhasspypiper_2024}, and sent it back to the client. The client would then use the Web Audio API and Tone.js \cite{noauthor_tonejs_nodate} to make the binaural and pitch adjustments respectively. 

On the client-side, some of our most important architectural considerations are as follows:

\textbf{Custom Input Fields} We developed a custom input field and overrode the built-in screen reader feedback. Although normally discouraged, this was necessary for a few reasons:

First, we needed custom hooks to determine exactly when the cursor moved, to override the default behavior upon falling off either edge of the input field. This let us seamlessly navigate across our tree-based space. The built-in HTML input element was much less consistent, and added extraneous and confusing announcements such as "end of text." 

Second, it allows the user to add LaTeX symbols, without using code. E.g., the user could input a symbol such as "$\delta$", and it would function as a single character and read as "delta," while preserving the LaTeX code "$\backslash$delta" under the hood. 

Third, and perhaps most importantly, it allowed us to modify and customize the speech output audio, which was critical for spatial sound. For example, consider an input field with the following text, with the "|" representing the caret: "a|bcde". If the user were to scan, character by character, to the end, the user would notice that the audio moves from left to right, with "e" being played fully on the right speaker. Thus, we hijack every possible access point, and make the entire user experience spatial. 

\textbf{Navigation Algorithm} We wanted our traversal order to grant a linear navigation style, and follow the same reading order as MathSpeak. Since our tree used leaf nodes as input fields, and internal nodes as operators, we used a pre-order traversal. We also exploited the tree-based structure to improve rendering performance through memoization. 

\textbf{Keyboard Spatial Navigation Algorithm} This provides a muscle memory, tactile interface while reusing the user's own keyboard. We take the middle four rows of the keyboard, and, based on the user's keyboard layout, find the physical relative position of each key. A high level example is as follows: Take the equation $\frac{a}{b}+\frac{c}{d}+\frac{e}{f}$. Notice that it is essentially a grid, with two rows and three columns. If the user presses a key in the top right of the keyboard, such as "p," the input field containing "$e$" will be highlighted. In addition, depending on whether the user is in "row mode" or "column mode," the editor will also play a series of notes representing the number of elements in the current row/column, as well as another note to indicate which item is currently selected. 

We also use our table representation to allow for arrow key navigation. We keep track of a virtual cursor, and move in exactly one dimension each time, clamping to the nearest input element. 

\textbf{Earcon Considerations} All earcons were inspired by our background in music. For example, the "parentheses" earcons are each a series of tones, one going up and the other going down. Combined with strong transients, we replicated the feeling of opening/closing a book.

\section{Future Directions}
\label{sec:discussion}

Based on our co-design sessions, preliminary results look promising. Thus, we are in the process of conducting a user study to gain feedback and produce an even more intuitive tool. Later, our work could even be extended to construct a full-fledged mathematical working environment, solving a major gap in the space \cite{stoger_accessing_2015}.

\begin{acks}
  We express our highest gratitude for all the support that made this research possible. This project was undertaken independently and received no external funding, and was a personal endeavor driven by our own passion, curiosity, and desire to see change happen. Special thanks to Nikki Kauffman and Tom Rix from Red Hat, Inc. for their undying support and encouragement. 

This is my (Kenneth Ge's) first publication, and deeply meaningful in what it represents. I am abundantly grateful for all of the support, and fearfully excited about the journey ahead.
\end{acks}

\bibliographystyle{ACM-Reference-Format}
\bibliography{references/nl+latex,references/a11y_framework}

\end{document}